\documentclass[sn-mathphys,Numbered]{sn-jnl}% Math and Physical Sciences Reference Style
\usepackage{graphicx}%
\usepackage{multirow}%
\usepackage{amsmath,amssymb,amsfonts}%
\usepackage{amsthm}%
\usepackage{mathrsfs}%
\usepackage[title]{appendix}%
\usepackage{xcolor}%
\usepackage{textcomp}%
\usepackage{manyfoot}%
\usepackage{booktabs}%
\usepackage{algorithm}%
\usepackage{algorithmicx}%
\usepackage{algpseudocode}%
\usepackage{listings}%
\usepackage[colorinlistoftodos,textwidth=50,textsize=tiny]{todonotes}

\usepackage{soul}
\usepackage{ulem}

\begin{document}

\newcommand\green[1]{\textcolor{green}{#1}} % EB comments and modifications
\setstcolor{green} % Just for EB crossed text
\newcommand\purple[1]{\textcolor{purple}{#1}} % MIR comments and modifications
\setstcolor{purple} % Just for MIR crossed text
\newcommand{\CA}[1]{{\color{teal}#1}} % CA comments and modifications
\setstcolor{teal} % Just for CA crossed text
% \usepackage{hyperref}

% \title[Article Title]{Beyond Classical and Data‐Driven Approaches: A Physics‐Informed Automatic Differentiation Scheme for Bragg CDI Phase Retrieval}

% \title[Article Title]{ConvexAD: A geometrically regularized differentiable model for 
% high-strain and low-signal BCDI phase retrieval.}

\title[Article Title]{ConvexAD: differentiable phase retrieval with geometric 
regularization for high-strain, low-signal Bragg coherent diffraction imaging}

% \green{ Maybe to distinguish it from other DL BCDI papers, you can give a name to your method (like autophaseNN from Cherukara, PID3Net, etc... Also I would chang the name to include something about reciprocal phase. So the title already poitn the difference between your technique and CHerukara, Robinson's papers}

\author[1,*]{\fnm{Matteo} \sur{Masto}}
\author[2,]{\fnm{Ewen} \sur{Bellec}}
\author[1,]{\fnm{\'Etienne} \sur{Polack}}
\author[2,]{\fnm{Jiangtao} \sur{Zhao}}
\author[2,3]{\fnm{Vincent} \sur{Favre-Nicolin}}
\author[2]{\fnm{Steven} \sur{Leake}}
\author[2]{\fnm{Tobias} \sur{Schülli}}
\author[1]{\fnm{Marie-Ingrid} \sur{Richard}}
\email{matteo.masto@esrf.fr}

\affil[1]{\orgname{CEA Grenoble, IRIG, MEM, NRX}, \orgaddress{\street{17 rue des Martyrs}, \city{Grenoble}, \country{France}}}

\affil[2]{\orgdiv{ESRF}, \orgname{The European Synchrotron}, \orgaddress{\street{Av. de Martyrs, 71}, \city{Grenoble}, \country{France}}}

\affil[3]{\orgname{Univ. Grenoble - Alpes}, \orgaddress{\city{Grenoble}, \country{France}}}

%\affil[2]{\orgname{CNRS, Grenoble INP, Institut Néel}, \orgaddress{\street{25 rue des Martyrs}, \city{Grenoble}, \country{France}}}


\maketitle

Bragg Coherent Diffraction Imaging (BCDI) relies on phase retrieval (PR), a non-convex, ill-posed inverse problem whose conditioning is degraded by factors like noise, partial coherence, and strain or crystallographic defects in the imaged nanocrystal. Conventional approaches, including alternating projections and deep learning methods, represent the object on a dense voxel grid, which is prone to amplitude-phase mixing and becomes increasingly ill-posed under these factors, often preventing reliable three-dimensional mapping of the particle's shape and strain. Here we present ConvexAD, a gradient descent PR model that parametrizes the object shape as a convex polyhedron, built by the intersection of half-spaces and  enforces flat amplitude within it by construction. This geometric regularization eliminates amplitude-phase mixing and reduces the number of trainable parameters, hence facilitating the convergence of the PR. Implemented in JAX for native GPU acceleration and forward-model flexibility, ConvexAD is validated on simulated and experimental BCDI data, where it outperforms conventional and deep learning assisted PR at high strain, low signal, and reduced coherence, extending the conditions under which quantitative BCDI reconstruction is achievable.




\section{Introduction}\label{sec1}
The development of Bragg Coherent Diffraction Imaging (BCDI) has enabled new opportunities for the experimental investigation of crystalline nanoparticles~\cite{Robinson2009CoherentNanoscale, PhysRevLett.87.195505}. This lensless technique performed at synchrotron and free-electron-laser facilities exploits the improved coherence of $4^{th}$ generation X-ray light sources to produce three-dimensional (3D) images of single nano-crystals' shape and internal lattice deformations with high spatial resolution (down to $\sim$ 10 nm), strain sensitivity ($\sim$ $10^{-4}$) and sensitivity to crystallographic defects~\cite{labat2015inversion,PhysRevMaterials.4.013801, comby2025combining,Lim2021ADiffraction}. BCDI has been successfully performed to study nanoparticles with size ranging from 20 nm~\cite{Richard:te5091} to 1 $\mu$m in combination with different sample environments such as the monitoring of nanoparticle catalytic activity \cite{grimes2024capturing} during electro-chemical reactions \cite{Atlan2023,Vicente2026, atlan2025probing,olson2025kinetics}, low or high temperature \cite{yang2025low, chatelier2024unveiling} and pressure conditions \cite{Yang2013,zakaria2026strain}.

In a typical BCDI experiment, a monochromatic coherent X-ray beam fully illuminates the sample, which is mounted on a goniometer. The detector, positioned in the far-field regime, captures 2D slices of the Bragg reflection diffraction pattern. During the so-called rocking curve, the particle is rotated in order to collect the full 3D diffracted BCDI pattern.
%\purple{Note that the 3D BCDI diffraction pattern can also be obtained by scanning the X-ray energy around a Bragg peak \cite{cornelius_three-dimensional_2011}.}%  
The full 3D pattern, in the kinematic approximation, is the squared modulus of the Fourier transform of a complex object which is composed of an amplitude, representing the shape and electronic density of the sample, and a phase, which encodes the projection of the lattice displacements onto the Bragg scattering vector \cite{Godard:te5072}. 
The squared modulus operation in the detector space eliminates the phase information associated with the diffracted signal thus preventing the retrieval of the object shape and lattice displacements via inverse Fourier transform. This is the so called ``phase problem''.
In absence of a closed-form mathematical expression mapping the amplitude and the phase of the diffraction pattern, the phase retrieval (PR) is attained using iterative algorithms \cite{Favre-Nicolin2020PyNX:Operators, Gerchberg1972APA}.

Since the early developments of CDI, alternating projection (AP) algorithms such as Error Reduction (ER), Hybrid Input-Output (HIO) \cite{Fienup:78} and Relaxed Average Alternating Reflections (RAAR) \cite{Luke_2005} have been mostly employed for the PR of BCDI datasets \cite{Marchesini2007ARetrieval}. This class of iterative algorithms operates by projecting the estimate back and forth between object and Fourier space, applying a constraint in each: a compact support in object space, and the measured diffraction signal in Fourier space. The solution is found when the estimate converges to the intersection of the two spaces' constraints. 
In BCDI, the convergence of PR is found to be hindered by low photon counts \cite{cdi_noise2017}, partial coherence of the probe \cite{clark_high-resolution_2012}, and most importantly by the presence of large crystal deformations such as strain $> 0.1\%$ \cite{minkevich_inversion_2007} .The signature of the latter is a severe distortion of the diffraction pattern and a large phase range $>2\pi$ in the complex object. Though a rigorous and agreed explanation of what precisely afflicts PR is yet to be found, it is evident that the problem correlates with the magnitude of the object's phase range. A reasonable hypothesis is that large phase ranges exceed the $[0,2\pi)$ interval, accumulating multiple wraps and hence introducing points that could lead to the formation of local minima in the PR solution landscape. Another   
common choice to overcome this problem is to run the PR algorithms multiple times, starting from different initializations of the estimate, to increase the chances of beginning in a region of the search space close to the solution. However, in addition to being both time and power-consuming, this method does not always succeed in reaching the solution.
 
Over the years, several approaches have been proposed to improve the PR for the ``strong-phase'' (also known as ``high-strain'') case. Common to these approaches is a more regularized formalization of the PR problem with external constraints based on prior knowledge or learned priors. The first category includes Bragg Coherent Modulation Imaging \cite{zhao2025bragg,Zhao}, which modulates the diffracted beam with a known 2D modulator, and the Bragg Ptychography technique, which exploits multiple illuminations with overlaps \cite{hruszkewycz2017high,takahashi2013bragg,li2021revealing}. Prior knowledge of the particle shape from a low-strain state can be enforced during the reconstruction of the same particle at high-strain states \cite{wang2020algorithm}. Alternatively, the expected flat electron density allows for a compressed sensing approach, which regularizes the PR by minimizing the amount of information required to express the object's amplitude \cite{NewtonStrain}.
% \st{More recently, deep learning approaches have enabled to leverage a learned form of prior knowledge to attain better PR results.} 
More recently, machine learning \cite {pelzer2021removal, bellec2026machine} and deep learning approaches \cite{Masto:yr5131} have been used for data pre-processing and PR by leveraging a learned form of prior knowledge to achieve faster or better results \cite{Cherukara2018Real-timeNetworks, Yao2022AutoPhaseNN:Imaging, Wu:cw5029, Masto2026}.

% \green{(I think there need to be a separate sentence about the deep-learning model for high-strain you did. The referee will ask about it. Should present it as "its good but there're still limitations" leading to AD. Or otherwise you talk about the comparison in the discussion of your last figure, as you prefer. Careful not to trash it too much. For example in the discussion, you can say that your DL model is still much faster for a prediction than AD.)}

From a different perspective, the PR can be seen as a minimization problem, suitable for gradient descent methods. A metric defining the discrepancy between the observed data and the forward model is minimized by updating the variables defining the amplitude and phase of the unknown object. This approach was first considered by Jurling and Fienup \cite{Jurling:14}, but then abandoned due to the lack of fast and efficient gradient computations for large image data. Today, machine learning libraries offer easy use of analytical derivatives for any differentiable operator through automatic differentiation, often natively accelerated on GPUs. This great flexibility, coupled with computational efficiency opens a new paradigm in the field of inverse problems, which has already been explored in the Coherent Diffraction Imaging community \cite{kandel2019using, Maddali2023, Wu2025}. 

Here we present ConvexAD, a gradient descent approach for the BCDI phase problem that exploits the seamless integration of complex forward models within automatic differentiation (AD) to implement constraints based on the prior assumption of compactness and homogeneity of the support of the measured object. By enforcing this regularization, the method successfully addresses most of the main limitations of conventional PR, enabling reliable reconstruction of highly strained samples while remaining robust to low-signal and partial-coherence conditions. Representative examples and quantitative comparisons with conventional PR are discussed in the paper.

% \green{Something missing for me is the "facet biais" that you use. It would be nice to talk a little about it before starting the model description where faceting is visible in Figure 1. Bon actually you talk about it later so maybe it's ok.}

\begin{figure}[H]
    \centering
    \includegraphics[width=\textwidth]{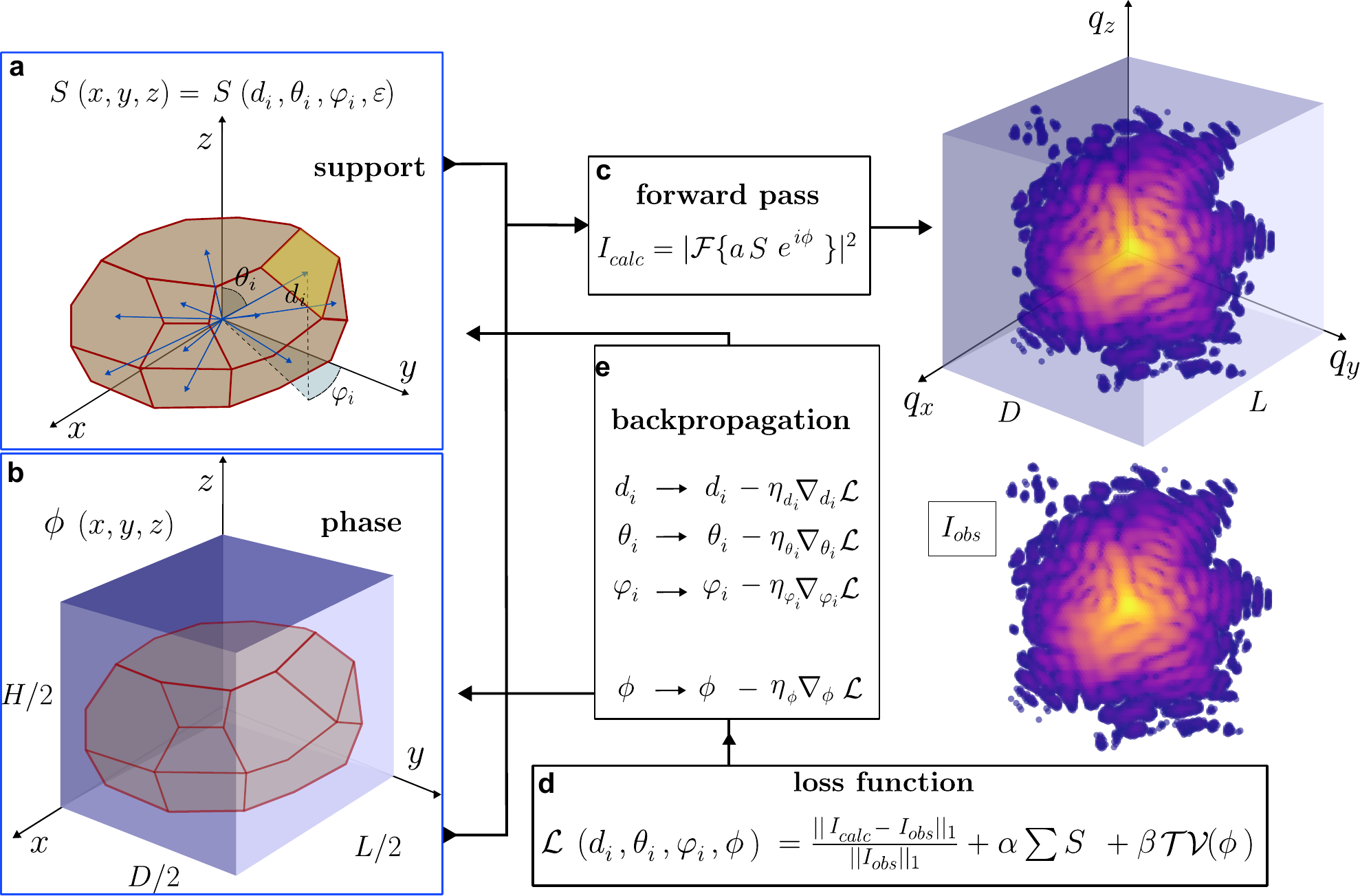} 
    \caption{\textbf{ConvexAD schematic.} \textbf{a)} The object support is parametrized with the half-spaces method. $N$ vectors are expressed in spherical coordinates $(d_i, \theta_i, \varphi_i)$ where $d_i$ is the length of the \textit{i-th} vector measured from the center of the 3D array. At the end of each vector a plane is built perpendicular to it. The intersection of all the $N$ planes is calculated and the support $S(x,y,z)$ is defined by the convex region produced by the planes intersection. A sigmoid function of steepness $\varepsilon$ filters the binary support, making it differentiable and introducing a spatial resolution variable.  \textbf{b)} The phase $\phi(x,y,z)$ is initialized on a 3D grid of half the size of the diffraction pattern window to account for the oversampling condition. \textbf{c)} The BCDI forward model represented by the squared modulus of the Fourier transform of the complex object built from $(a, S, \phi) $ is calculated, where the amplitude $a$ is a scalar value obtained from Parseval's theorem. \textbf{d)} The loss function compares the calculated intensity with the observed data. In order from the left there are the normalized mean absolute error (MAE) as data fidelity term, the penalty on the support size, tuned by the coefficient $\alpha$, and the total variation (TV) on the phase, tuned by the coefficient $\beta$, as regularization terms. \textbf{e)} Each trainable variable is updated during the backpropagation along the steepest gradient direction, multiplied by the adaptive learning rate produced by the ADAM optimizer.
    % \green{(1) caption is too long. Details should be in main text. It's ok if the caption doesn't fully explain the figure. (2) Is k the epoch ? If possible I would remove it. it would make all formulas easier to read. for the backpropagation you can use $d \xrightarrow{} d-... etc,$ (3) Is $i$ the plane index ? If so, why is there a $i$ on $\phi_i$ ? (4) Is $a$ trainable ? If so, it should be in the arguments of the loss in d) (5) why is there a different training rate for each argument in the backpropagation ?(6) Should we use $\nabla_\theta$ instead of $\nabla L (\theta)$, depend what you think is more readable. (7)Why is there a $\gamma$ in the loss in d), isn't it redundant with argument $a$? (8) Use \textbf{a} instead of (a) like the other figures (9) also use symbols for calculated intensity and observed intensity for reference in the main text.}
    }    
    \label{fig:ad_scheme} 
\end{figure}

\section{ConvexAD: a convex, differentiable support parametrisation for BCDI}\label{sec1b}

The three-dimensional object is reconstructed from the measured coherent
diffraction pattern using a gradient descent optimization scheme, referred
to hereafter as ConvexAD, in which the object support, and
phase are refined jointly through backpropagation of the forward diffraction model. 
Figure \ref{fig:ad_scheme} explains the approach. For a given
diffraction pattern, ConvexAD initializes a batch of independent complex
3D arrays that are optimized in parallel with GPU acceleration during
PR. Before optimization, each complex object is built
from a trainable support mask and a trainable phase array, defined by
default on a grid half the size of the input diffraction pattern,
consistent with the assumption that the oversampling condition is already
fulfilled; a custom grid shape can also be specified to accommodate
sub-optimal oversampling conditions. Central to the model is the
parametrization of the object support.

In BCDI, the amplitude of the reconstructed complex object represents the
electron density of the crystal, which, in the general case of a
single-material particle, is expected to be homogeneous within the
object support, apart from localized holes caused by the strong strain
induced by crystallographic defects. The flatness of the amplitude is commonly
taken as a signature of good PR convergence, to the point that the
mean-to-max metric has been proposed as a criterion for ranking
reconstruction quality \cite{frisch2023unraveling, grimes2024capturing}.
This prior knowledge can be encoded directly as a regularization,
enforcing a homogeneous amplitude by construction in the object
formulation. This approach was first proposed by Minkevich \textit{et al.}
to facilitate the PR of highly strained BCDI patterns \cite{Minkevich2008}.
Regularization of this type is also well suited to avoiding the
detrimental phase-amplitude mixing that leads to an incorrect object
support, a common issue during PR of highly strained particles
\cite{PhysRevB.83.224109, Zhao:te5119}.

Rather than representing the support as an independent binary voxel grid, as is common in standard PR algorithms, we assume a constant amplitude value inside a compact shell, which allows the optimization to focus solely on the object surface, reducing the number of unknowns and
restricting the search space accordingly. This number is further reduced
when the surface itself is parametrized as the geometric intersection of
a set of planes, defining a compact, convex, and faceted surface. While
this representation cannot directly accommodate concave morphologies or
disconnected domains, it is well-suited to the faceted crystalline
nanoparticles commonly investigated with BCDI. By embedding this
geometric prior into the reconstruction model, the set of admissible
solutions is substantially reduced, providing strong regularization and
improved optimization stability.

As shown in Fig.~\ref{fig:ad_scheme}\textbf{a}, ConvexAD uses the assumption of a homogeneous amplitude inside the
object support by constructing a convex hull using the half-spaces method. $N$ vectors $\mathbf{n}_i$,
expressed in spherical coordinates $(d_i, \theta_i, \varphi_i)$ for
$i = 1, \dots, N$, originate from the center of the reconstruction grid:
\begin{equation}
    \mathbf{n}_i \;=\;
    \begin{pmatrix}
    \sin\theta_i\cos\varphi_i \\[6pt]
    \sin\theta_i\sin\varphi_i \\[6pt]
    \cos\theta_i
    \end{pmatrix},
    \label{eq:normal_vector}
\end{equation}
where $d_i$ denotes the distance from the center to the $i$-th facet and
$(\theta_i, \varphi_i)$ define its orientation. At the tip of each
vector, a half-space is defined by the plane perpendicular to it, and the
object support $S(x,y,z)$ is obtained as the convex intersection of the
$N$ resulting half-spaces:
\begin{equation}
    \mathcal{S}(\mathbf{x}) \;=\; \prod_{i=1}^{N}
    \sigma\!\left(\frac{d_i - \mathbf{n}_i \cdot \mathbf{x}}{\varepsilon}\right),
    \qquad \sigma(t) = \frac{1}{1+e^{-t}}.
    \label{eq:convex_hull}
\end{equation}
The trainable variables are therefore the tuple $(d_i, \theta_i, \varphi_i)$
for each of the $N$ vectors, with $N$ provided by the user as an external
parameter, which typically ranges from a few tens for faceted particles to a
few hundred for more rounded particles. 
To render this construction differentiable, and therefore trainable, the hard intersection is relaxed into the smooth sigmoid function  $\sigma$  in Eq. ~\ref{eq:convex_hull}. This also introduces a controllable spatial resolution at the object boundary, set by a user-defined steepness parameter $\varepsilon$ 
that converges to a binary mask in the limit of small $\varepsilon$. 
% Although rarely necessary, $\varepsilon$ can also be
% updated during optimization, starting from a loose support (large
% $\varepsilon$) and progressively tightening it. 
This low-parameter convex description constrains the reconstructed morphology to physically
plausible convex shapes and reduces the number of free parameters by
orders of magnitude compared to a voxel-wise support, while remaining
fully compatible with gradient-based optimization: for a 3D diffraction
pattern of size $(H, L, D)$, the number of unknowns is reduced from
$H/2 \times L/2 \times D/2$ voxels (e.g. 125000 parameters for a $100\times100\times100$ pixels dataset) to $3 \times N$ variables (e.g. 300 parameteres for 100 half-spaces), narrowing
the solution space and facilitating PR. 

The phase $\phi(x,y,z)$ (Fig.~\ref{fig:ad_scheme}\textbf{b}), in
contrast, is left unconstrained: ConvexAD automatically initializes a
trainable 3D array with random values drawn from a uniform distribution,
defined on a grid half the size of the diffraction pattern window along
each dimension, in accordance with the oversampling condition required
for PR. In the forward pass, the model computes the
diffracted intensity from the complex object, which is padded with zeros to
match the size of the observed data (Fig.~\ref{fig:ad_scheme}\textbf{c}).
At each iteration, the complex object is constructed as $a\,S\,e^{i\phi}$,
where the scalar amplitude $a$ is fixed by Parseval's theorem to
conserve the total measured signal energy,
\begin{equation}
    a = \sqrt{\frac{\sum I_{\mathrm{obs}}}{\sum \mathcal{S}^2}},
    \label{eq:amplitude}
\end{equation}
and the corresponding diffraction intensity is calculated as
\begin{equation}
    I_{\mathrm{calc}} = \left| \mathcal{F}\{a\,S\,e^{i\phi}\} \right|^2 .
\end{equation}

The reconstruction proceeds by minimizing the loss function
\begin{equation}
    \mathcal{L}(d_i, \theta_i, \varphi_i, \phi) =
    \frac{\| I_{\mathrm{calc}} - I_{\mathrm{obs}} \|_1}{\| I_{\mathrm{obs}} \|_1}
    + \alpha \sum S
    + \beta\, \mathrm{}{TV}(\phi),
\end{equation}
where the first term is the normalized mean absolute error (MAE) between
the calculated and observed intensities, serving as the data fidelity
term and which has consistently outperformed alternative metrics in
our tests; the second term penalizes the total support size with weight
$\alpha$, encouraging convergence toward the smallest support that
minimizes the loss and preventing spurious growth of the reconstructed
object during optimization \cite{Favre-Nicolin2020}; and the third term
applies total variation (TV) regularization to the phase within the current support estimate with weight
$\beta$, penalizing unphysical phase variations while preserving genuine discontinuities, such as those associated with lattice defects.
(Fig.~\ref{fig:ad_scheme}\textbf{d}).

All trainable parameters, the support vectors $(d_i, \theta_i,
\varphi_i)$ for $i = 1, \dots, N$ and the phase field $\phi$, are
updated at each iteration by descending the gradient of $\mathcal{L}$,
with gradients computed automatically by JAX
\cite{jax2018github} and backpropagated through the forward model
(Fig.~\ref{fig:ad_scheme}\textbf{e}):
\begin{equation}
\begin{aligned}
    d_i      &\leftarrow d_i - \eta_{d_i} \nabla_{d_i} \mathcal{L}, \\
    \theta_i &\leftarrow \theta_i - \eta_{\theta_i} \nabla_{\theta_i} \mathcal{L}, \\
    \varphi_i &\leftarrow \varphi_i - \eta_{\varphi_i} \nabla_{\varphi_i} \mathcal{L}, \\
    \phi     &\leftarrow \phi - \eta_{\phi} \nabla_{\phi} \mathcal{L}.
\end{aligned}
\end{equation}
The per-parameter step sizes $\eta_{d_i}$, $\eta_{\theta_i}$, $\eta_{\varphi_i}$, and
$\eta_{\phi}$ are adapted automatically at each iteration using the ADAM
optimizer \cite{kingma2017adam}, which accounts for the differing scales
and sensitivities of the support and phase variables and improves both
the convergence rate and robustness of the reconstruction.

The optimization typically runs for several hundred iterations, with
all complex objects in the batch optimized in parallel; the object
achieving the lowest MAE score at the end of the run is selected as the
best solution. During a given PR run, the optimization can be paused to
inspect the current result and, if necessary, adjust the strength of the
regularization terms, the learning rate, or the resolution parameter
$\varepsilon$, before resuming until the loss reaches a satisfactory
value. From experience, a MAE value around 0.1, in the absence of other
penalty terms, is generally sufficient to achieve satisfactory
reconstructions with high confidence; under low signal-to-noise ratio
(SNR) conditions, however, this threshold is typically higher, as the
model cannot fully capture the effect of noise on the amplitude
homogeneity of the object. Finally, to validate and refine the ConvexAD
reconstruction, a few cycles of Error Reduction (ER) \cite{Fienup:78} are run at the end
of the optimization: when convergence is good, ER confirms the result;
when convergence is poor, it diverges from it, revealing an incorrect
solution.

\section{Results}\label{sec2}

Conventional PR has been demonstrated on a wide range of datasets but still fails for highly strained particles or for noisy, low-intensity data. Here, we apply ConvexAD to both noisy and highly strained datasets to demonstrate its robustness on these challenging cases.

% \green{(I think this paragraph is too much details that should rather be for the users of the code maybe as a text on github. I would just keep discussion on a rouh number of epochs + ER without talking about bad reconstructions such as : "The model is trained for ... epochs. From experience, a MAE value below 0.1 is, with high confidence, sufficient to achieve satisfactory reconstructions. However, under low signal-to-noise-ratio (SNR) conditions, this value is typically higher, as the model cannot fully capture the effect of noise on the amplitude homogeneity of the object. In order to validate and fine-tune the ConvexAd reconstruction, a few cycles of Error reduction (ER) are run at the end." or something like that.)}

\subsection{Case of noise and partial coherence in data: simulation and experiment}

Robustness to low SNR is a practical requirement for BCDI, as several experimental constraints impose reduced photon flux or acquisition time. In operando electrochemistry and battery studies, limiting the dose is necessary to avoid beam-induced perturbation of the reaction or artificial degradation of the material \cite{Björling2023, atlan2025probing, Jousseaume2023, Black2024}. Similarly, in some cases, resolving strain evolution as it is induced requires low incident flux to avoid perturbing the process under study \cite{SIMONNE2026116940}, while time-resolved BCDI inherently limits the photon budget per frame to achieve sufficient temporal resolution \cite{grimes2024capturing}. Reconstruction algorithms that remain reliable under low-photon-counts conditions are therefore essential for extending BCDI to beam-sensitive and fast-evolving systems. 

\begin{figure}[H]
    \centering
    \includegraphics[width=\textwidth]{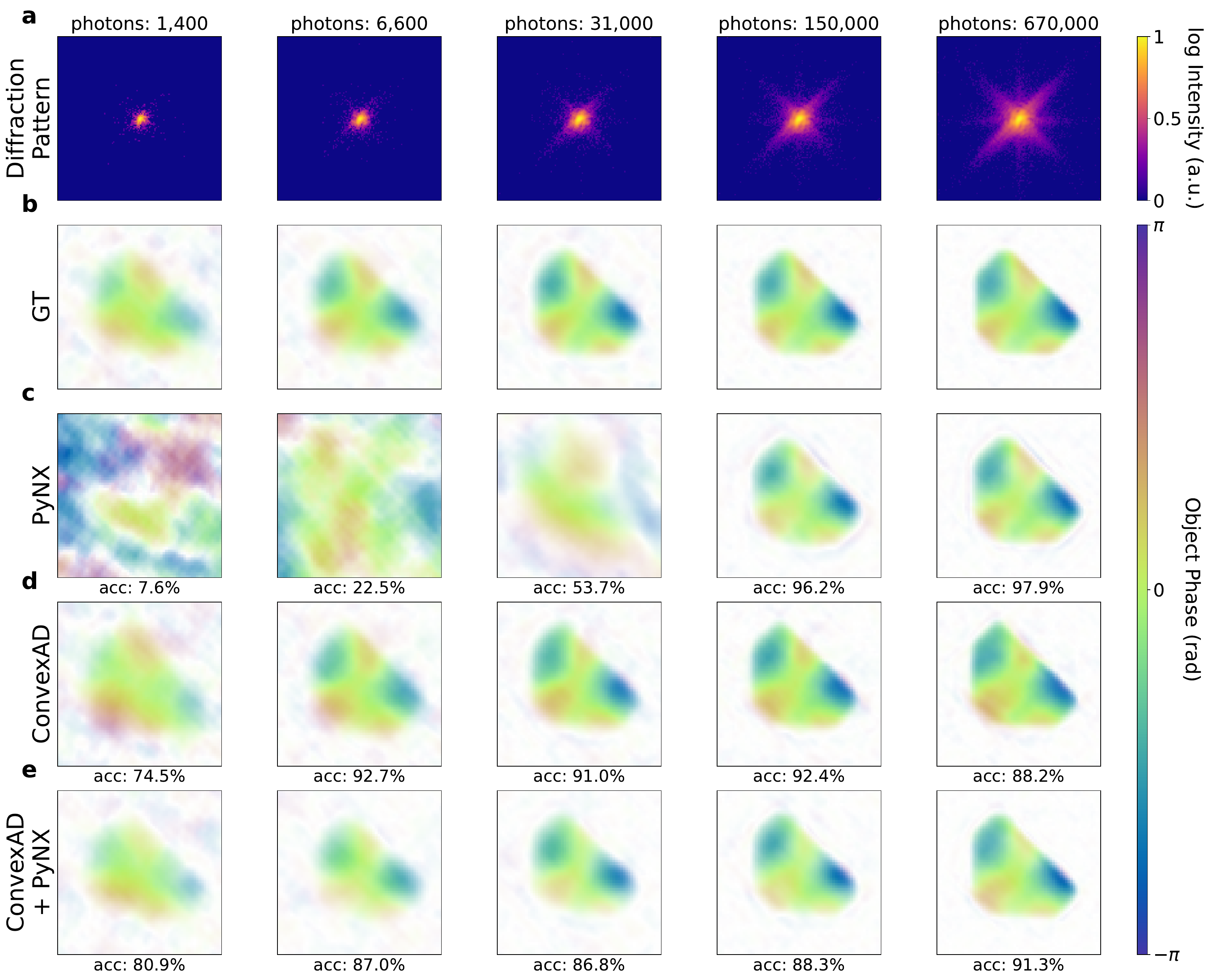} 
    \caption{\textbf{Results for low SNR conditions.} \textbf{a} Projection along first axis of simulated diffraction patterns for different levels of Poisson noise. The data is shown in logarithmic scale and normalized between 0 and 1 for better visualization and the total amount of photons is reported above each image. \textbf{b} Central slice of the object phase obtained from the inverse Fourier transform of the complex noisy diffraction pattern. \textbf{c} Results obtained with ``PyNX'' conventional iterative algorithms. \textbf{d} Results obtained from ConvexAD. \textbf{e} Results after ER refinement. In this case the results do not always improve their accuracy.  }
    \label{fig:noise} 
\end{figure}

Figure \ref{fig:noise} illustrates the impact of Poisson noise on PR by comparing reconstructions across a range of photon counts (from $1\times10^3$ to $6\times10^5$), spanning low to high SNR conditions. For each noise level, the simulated diffraction pattern is shown alongside the ground-truth object phase and the phase retrieved by ConvexAD, ConvexAD with ER refinement, and conventional iterative algorithms from the PyNX package \cite{Favre-Nicolin2020PyNX:Operators}, with reconstruction accuracy reported for each case. The accuracy is computed through the cross-correlation with the equation as follows:

\begin{equation}
\mathrm{Accuracy}(\mathbf{\Psi},\mathbf{\Psi_{\text{GT}}}) =
\left[
\frac{\left|\sum_{i=1}^{N} \Psi_i^{*} \Psi_{\text{GT},i}\right|}
{\sqrt{\left(\sum_{i=1}^{N} |\Psi_i|^2\right)
\left(\sum_{i=1}^{N} |\Psi_{\text{GT},i}|^2\right)}}
\right]
\times 100,
\label{eq:accuracy}
\end{equation}

where $\mathbf{\Psi}$ and $\mathbf{\Psi_{\text{GT}}}$ are the reconstructed and ground truth complex objects respectively. This metric has the advantage of taking into account the complex nature of the variables, naturally handling potential global phase offsets.Furthermore, the metric is not evaluated solely over the voxels inside the support, as this would necessitate selecting a threshold that is inherently arbitrary. Instead, the cross-correlation is computed over the entire complex-valued array, such that any non-zero amplitude outside the support lowers the correlation with the ground truth. 

Following the recipe described in the Supplementary Materials of Ref. \cite{Masto:yr5131}, the diffraction pattern of a 3D particle has been simulated under several noise conditions (see Fig. \ref{fig:noise}\textbf{a}). For each noisy diffraction pattern, the ``ground truth'' object has been obtained from the inverse Fourier transform of the complex diffracted signal, with the noisy amplitude replacing the noiseless one (Fig. \ref{fig:noise}\textbf{b}). The noisy ground truth object is in principle the best achievable from the given data and serves as a comparison with the results of ConvexAD and standard AP algorithms. The complete set of parameters used for each of the ``PyNX'', ``ConvexAD'', ``ConvexAD + ER'' reconstruction approaches is provided in the Methods section of the paper, 
% With ConvexAD, a batch of 32 estimates was optimized with $N=64$ half-space vectors and fixed $\varepsilon  = 0.4$ for 1000 iterations (Fig. \ref{fig:noise}\textbf{c}). ER refinement has been executed on each ConvexAD result (Fig. \ref{fig:noise}\textbf{d}). The PyNX package \cite{Favre-Nicolin2020PyNX:Operators} has been utilized for this step, for 300 iterations, with support threshold fixed at 0.25 and allowing only one voxel at the border to be updated every 50 cycles. Standard PR with AP algorithms has been performed with PyNX, following the scheme reported in the Methods section,
and the results are shown in Fig. \ref{fig:noise}\textbf{c-e}. 

The cross-correlation metric allows for a quantitative assessment of the reconstruction quality and was calculated between the ground truth and each reconstruction (see Eq. \ref{eq:accuracy}). At low photon counts, standard PyNX reconstructions degrade substantially, with accuracy as low as 7.6\% at 1,400 photons, while ConvexAD maintains high accuracy even under severe noise (74.5\% at 1,400 photons), demonstrating its robustness to low-signal conditions. ER refinement of the ConvexAD result does not consistently improve accuracy over ConvexAD alone and occasionally degrades it. We attribute this to the shrinkwrap support update, which likely over-shrinks the support when the signal is low. These results highlight the superiority of the ConvexAD method with respect to conventional AP algorithms: the regularization of the PR problem granted by the construction of a flat object amplitude guides the optimization towards the correct solution even when the raw signal is weak.

\begin{figure}[H]
    \centering
    \includegraphics[width=0.6\textwidth]{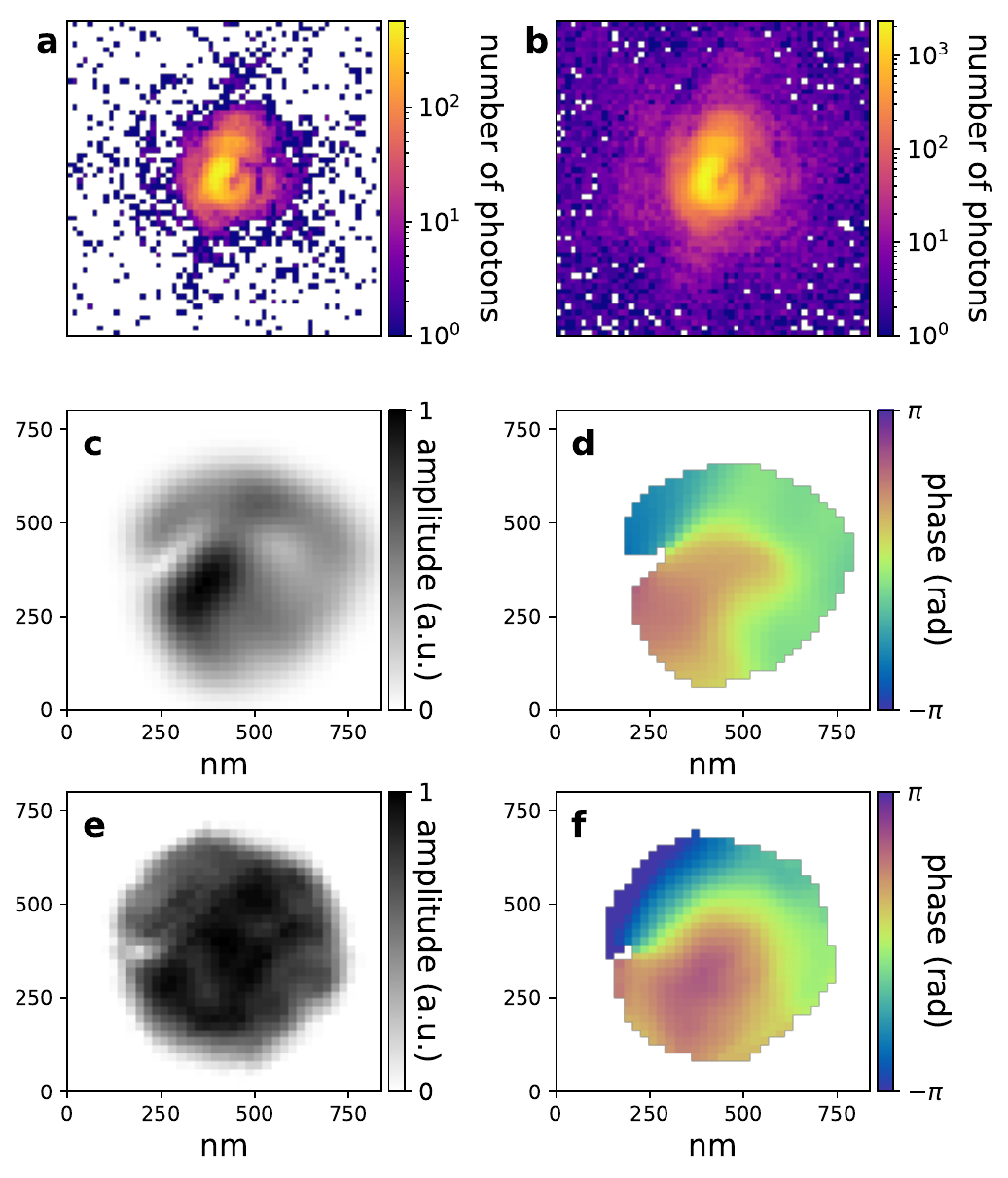} 
    \caption{ConvexAD robustness on low-photon and low-coherence BCDI. \textbf{a} BCDI data central slice and \textbf{b} projection. \textbf{c-d} Reconstructed object (amplitude and phase) using iterative algorithms. \textbf{e-f} Reconstructed object (amplitude and phase) using ConvexAD + ER refinement.} 
    \label{fig:bm02} 
\end{figure}

ConvexAD was tested on experimental data collected under low photon count conditions at the BM02/D2AM beamline of the ESRF synchrotron (see Methods section), which is served by a bending magnet source \cite{met9030352}. Besides the low photon counts caused by the lower flux of a bending magnet beamline compared to an insertion device one, the probing radiation has a reduced degree of coherence, which is known to negatively affect the PR \cite{Vartanyants_2001}. For the dataset reported here and shown in Fig.~\ref{fig:bm02}, the maximum intensity value is only 648 photons, with a total of 149737 photons summed over the array. This is low compared to typical BCDI measurements on Pt particles, which normally reach a maximum peak intensity of $10^4$--$10^5$ photons \cite{Carnis2019TowardsReconstructions}. Moreover, the degree of coherence is only 50\% in the vertical direction and 8\% in the horizontal direction, which makes reconstruction with standard iterative algorithms challenging, if not impossible, as illustrated in panels \textbf{c--d}.

Despite these constraints, ConvexAD successfully reconstructs the correct Winterbottom-shaped particle, as shown in panels \textbf{e-f}. The incorporation of facets and object amplitude homogeneity bias in ConvexAD leads to robustness against such experimental limitations.
    
\subsection{Case of highly strained and defective particles: simulation and experiment}
\begin{figure}[H]
    \centering
    \includegraphics[width=\textwidth]{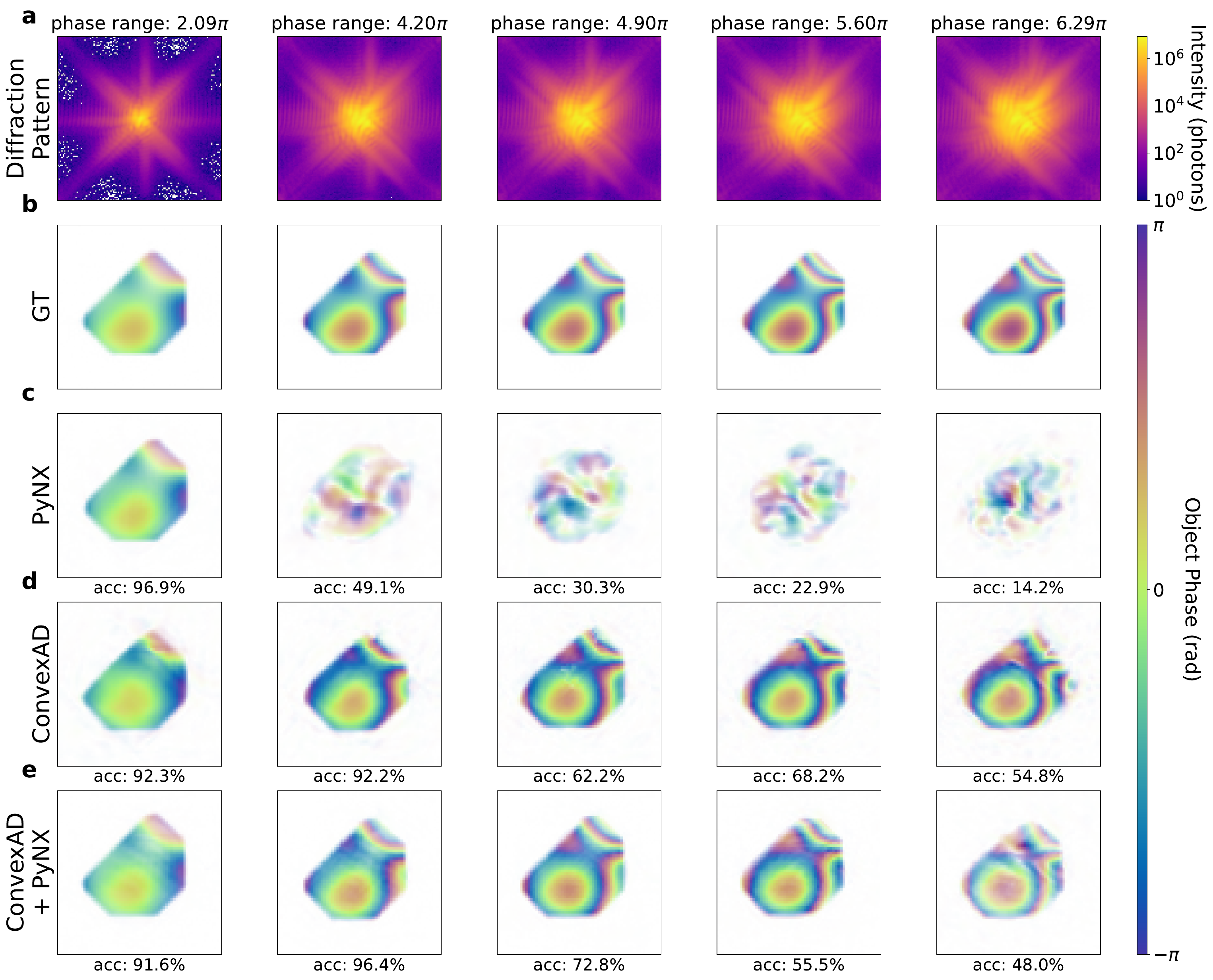} 
    \caption{\textbf{Results for simulated objects with large phase ranges.}
    \textbf{a} Projection along first axis of simulated diffraction patterns obtained from objects with increasing phase range. \textbf{b} Central slice of the ``Ground truth'' object phase. \textbf{c} Results obtained with ``PyNX'' conventional iterative algorithms. \textbf{d} Results obtained from ConvexAD. \textbf{e} Results after ER refinement. }
    \label{fig:strain} 
\end{figure}

As noted earlier, one of the major limitations of BCDI is the difficulty of reconstructing diffraction patterns from highly strained samples. This limitation can prevent BCDI from imaging many highly strained nanoparticles that are relevant to important scientific applications, highlighting the need for robust reconstruction algorithms capable of overcoming this challenge. ConvexAD was developed specifically to address this limitation. 
Similar tests have been conducted by varying the magnitude of the object phase range, to simulate highly strained crystals. A Gaussian-correlated random phase field \cite{Wu:cw5029} with progressively increasing amplitude was applied to the same 3D object shown in Fig. \ref{fig:noise}, and the corresponding diffraction patterns were calculated using a Fourier transform. The same level of Poisson noise was added to each diffraction pattern and the PR performed with ConvexAD, further refined with ER, as well as with the ``PyNX'' method adopted above. As expected, standard PR fails to reconstruct as the phase range increases, breaking down as the phase range exceeds a full phase wrap ($ > 2\pi$) as illustrated in Fig. \ref{fig:strain}\textbf{c}. In contrast, ConvexAD successfully retrieves the object up to phase ranges in the order of 3 wraps ($ > 6\pi$) 

\begin{figure}[H]
    \centering
    \includegraphics[width=\textwidth]{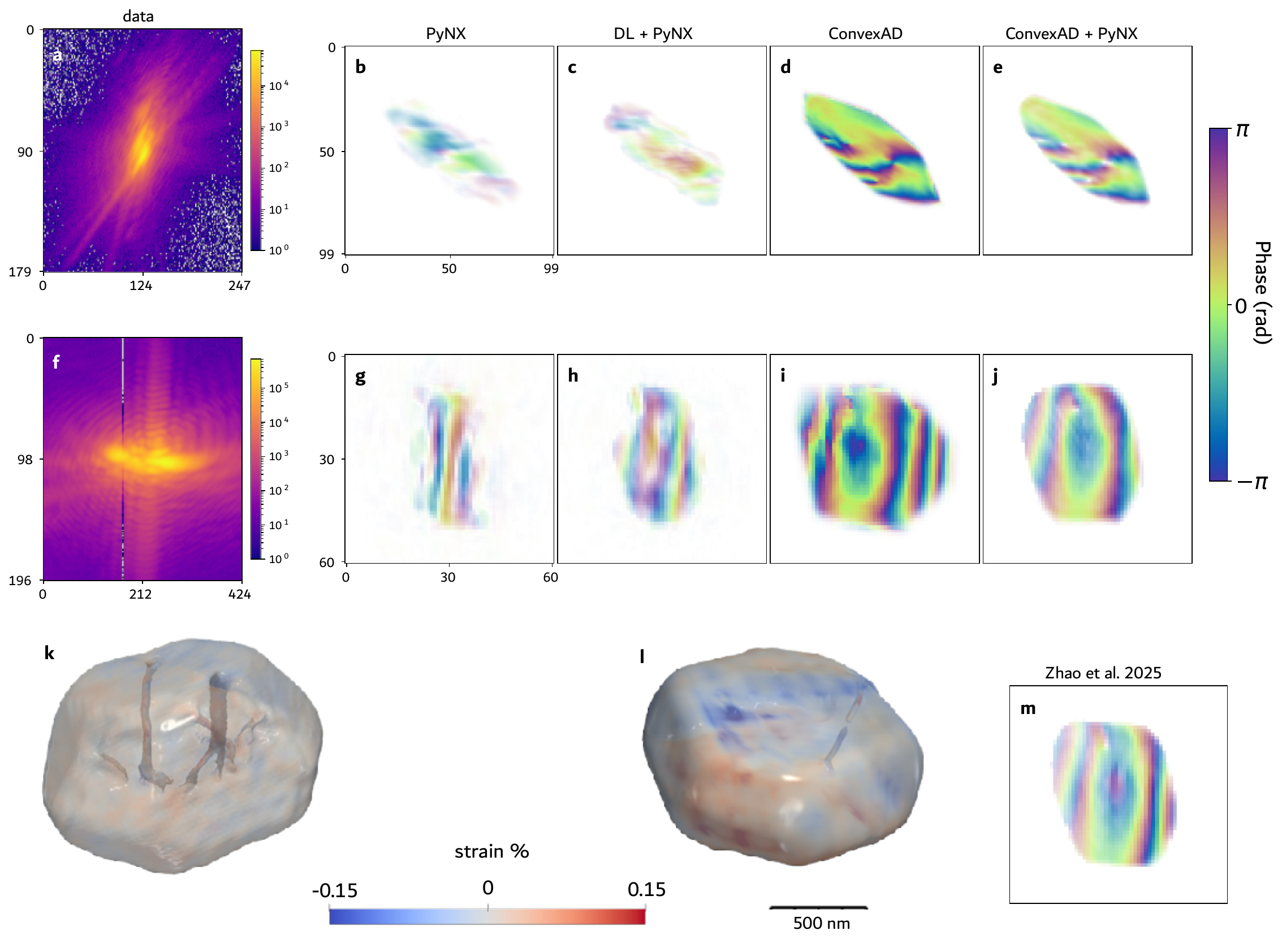} 
    \caption{\textbf{Results on experimental data from highly strained nanocrystal}. \textbf{a,f} Projection slice of the datasets. \textbf{b,g} Result of the PR with the ``PyNX'' method. \textbf{c,h} Result with the ``DL + PyNX'' method. \textbf{d,i} Result of ConvexAD. \textbf{e,j} Result after ER refinement of the ConvexAD solution. \textbf{k} 3D rendering of the Convex + PyNX solutions, with strain colormap after orthogonalization: \textbf{k} the particle with multiple dislocations and \textbf{l} the particle reported in \cite{zhao2025bragg}, the same slice of which is displayed in the panel \textbf{m}. }
    \label{fig:exp_recs} 
\end{figure}

Here we present two examples of experimental BCDI data, collected at the ID01 beamline of the ESRF (see Methods section), that could not be reconstructed with standard PR but were successfully phased with ConvexAD.
The first case corresponds to the BCDI diffraction pattern measured around the $\bar{1}13$ reflection of a Pt particle on an yttria-stabilized zirconia (YSZ) substrate. The particle presents multiple dislocations and the PR with standard AP algorithms as well as with Deep Learning (DL) \cite{Masto2026} based approaches is unsuccessful. The second dataset is the BCDI pattern around the 111 reflection of another Pt particle on YSZ substrate. Such a dataset is reported in the study conducted by Zhao et al. \cite{zhao2025bragg}, where the authors showed the impossibility to reconstruct it with standard PR without providing the support of the object, while convergence was found with the BCMI approach. Fig. \ref{fig:exp_recs} summarizes the results obtained for both datasets, reconstructed with different methods, reported in detail in the Methods section. Other examples are provided in section S1 of Supplementary Material.  

Both results obtained with ConvexAD converge to the expected Winterbottom-shaped particle, with clear facets and homogeneous amplitude within the support confirmed after ER refinement. It is worth noting that the presence of dislocations, usually signaled by phase vortex and relative amplitude dips in the defect core, does not break the flat amplitude assumption enforced by construction in ConvexAD. In fact, these dips do not indicate missing electronic density but result from the finite reciprocal-space bandwidth of the measurement: the large displacement gradients at dislocation cores generate high-q scattering that is only partially recorded, leading to apparent amplitude reductions in real space \cite{Ulvestad2017_Guided, maddali_differentiable_2022}. By enforcing a homogeneous amplitude, ConvexAD prevents the introduction of artificial density minima and instead explains the measured diffraction entirely through the phase field, enabling accurate recovery of complex displacement fields. For this reason ConvexAD approach proved to be beneficial in case of PR of particles with misoriented domains as well \cite{diao2020evolution, dzhigaev2021three} (see Supplementary Material, Section S2).

\section{Discussion}
ConvexAD improves the robustness and convergence of PR for high-strain BCDI datasets compared to standard AP algorithms by embedding an explicit geometric prior into a differentiable forward model. Conventional algorithms reconstruct amplitude and phase independently on every voxel of the grid, tied together only by loose constraints such as support and positivity, which makes the inverse problem severely ill-posed and prone to amplitude-phase cross-talk, especially for highly strained objects. ConvexAD instead parametrizes the object support as a convex polyhedron obtained via a half-space intersection and enforces a flat amplitude within it by construction, removing the amplitude-phase mixing problem and reducing the trainable parameters from the $O(N^3)$ voxels of a 3D grid to the handful of parameters describing the polyhedron facets.
The same reduction in degrees of freedom also benefits reconstructions from noisy or partially coherent data, since the optimization has less room to overfit to noise or absorb coherence loss into spurious amplitude or phase structure. 
% which in our tests gave more consistent reconstructions at low photon counts and under partial coherence than standard AP algorithms.

Even though ConvexAD cannot represent concave morphologies or spatially detached multi-domain objects, the convex-polytope formulation is precisely what keeps the parameter count low and the inverse problem well regularized, making it easier to reconstruct high-strain convex particles, which make up the majority of typical BCDI samples. 

Another benefit of formulating PR as explicit loss minimization is the possibility of using the loss score itself as a stopping criterion, applying the same tolerance across a dataset or across different methods. This is harder to achieve with alternating-projection algorithms, where a quality metric is typically only evaluated a posteriori, after the reconstruction has run through a fixed, predefined recipe of iterations, rather than being used to decide when to stop. Being able to set a common tolerance directly during the optimization makes comparisons between reconstructions, and between methods, more controlled and reproducible.

Finally, because ConvexAD is implemented as a differentiable JAX model, extending the forward model, for additional physical effects, alternative support parameterizations, or custom regularization terms, requires only modifying the forward computation itself, which is more flexible and easier to implement than with AP algorithms, thanks to the GPU-accelerated automatic differentiation provided by machine learning libraries.

\section{Methods}
The BCDI measurement shown in Figure \ref{fig:bm02} was performed at 8.33 keV on a Pt nanoparticle obtained by solid-state dewetting on a YSZ substrate \cite{data_bm02}. The sample was rotated on a range of 0.6$^\circ$ with angular step size 0.0075$^\circ$ with counting time at each point of 10 s. The degree of coherence was calculated following \cite{stoclet2025structural} with distance from the source of 27.2 meters and slits before the sample closed to 340$\mu$m.

For the BCDI measurement shown in Fig. \ref{fig:exp_recs}a the Pt particle on a YSZ substrate was measured at 9keV rotating the sample by an step size of 0.006$^\circ$ over a range of 1.2$^\circ$. The exposure time was 1 s per frame
\cite{data_mouad}.

For the BCDI measurement shown in Fig. \ref{fig:exp_recs}f the Pt particle on a YSZ substrate was measured at 9keV rotating the sample by an step size of 0.005$^\circ$ over a range of 1$^\circ$. The exposure time was 1 s per frame \cite{data4}.

Standard PR was performed using the PyNX software package. For each dataset, 100 independent reconstructions were generated, each comprising 400 iterations of HIO, 1000 iterations of RAAR, and 300 iterations of ER, with the support threshold varied randomly across runs between 0.1 and 0.25 and support updates restricted to border voxels every 30 cycles. The three reconstructions exhibiting the best Mean-to-Max metric were selected and combined by mode decomposition \cite{FreeLLK}. This procedure is referred to as ``PyNX'' in the text. 
For the first experimental dataset (Fig. \ref{fig:exp_recs}-\textbf{b}) it required about 7 minutes and 15 minutes for the second (Fig. \ref{fig:exp_recs}-\textbf{g}), owing to its larger size.
The deep-learning (DL) approach instead employed the convolutional neural network architecture proposed in \cite{Masto2026}, followed by a refinement stage of 300 ER cycles, run with PyNX, in which the support border was updated every 50 iterations. Updating only the border of the support is advised in case of poor initial estimate from the Patterson function, which is typical of the high-strain case. 

PR based on ConvexAD was performed in batches of 32 independent runs for simulated cases initializing 64 half-space vectors with $\varepsilon=0.6$, learning rate of 0.05 with exponential decay. 300 cycles of ER refinement were executed on ConvexAD results using PyNX and allowing the border of the support to be updated every 50 iterations. ADAM optimizer with a learning rate of 0.05 was used in all cases, both experimental and simulated.
Experimental datasets were reconstructed with ConvexAD using for the first dataset a batch of 16 independent runs, with $N = 64$, $\varepsilon = 0.5$ looping for 2000 iterations ($\sim$ 3 minutes) after which the best reconstruction yielded a normalized MAE of 0.14 (Fig. \ref{fig:exp_recs}-\textbf{d}).
For the second dataset (Fig. \ref{fig:exp_recs}-\textbf{i}), GPU memory limitations associated with its larger size necessitated prior binning of the data. The batch size was set to 64, $N = 128$ and $\varepsilon = 0.6$. The PR required 5000 iterations (approximately 50 minutes) to converge, yielding a final loss value of 0.08. Further refinement was performed with 300 cycles of ER using PyNX. (Fig. \ref{fig:exp_recs}-\textbf{e,j})

All PR methods were executed with GPU acceleration, provided in this study by an NVIDIA A40 with 48GB of memory.

\backmatter

\bmhead{Supplementary information} 
Supplementary information is available for this paper.
 
\section*{Declarations}

\begin{itemize}
\item Funding: We acknowledge funding from the ERC under the European Union's 
Horizon Europe programme, grant agreement No. 101201163.\\

\item Availability of data and materials: The datasets used during the current study available from the corresponding author on reasonable request. Mentioned experimental datasets are provided with corresponding DOI.
\item Code availability: The codes for this study are available and accessible via this link: 
\url{https://github.com/matteomasto/convexad}
\end{itemize}



\bibliography{sn-bibliography}% common bib file

@article{PhysRevB.83.224109,
  title = {Propagation uniqueness in three-dimensional coherent diffractive imaging},
  author = {Huang, Xiaojing and Harder, Ross and Xiong, Gang and Shi, Xiaowen and Robinson, Ian},
  journal = {Phys. Rev. B},
  volume = {83},
  issue = {22},
  pages = {224109},
  numpages = {4},
  year = {2011},
  month = {Jun},
  publisher = {American Physical Society},
  doi = {10.1103/PhysRevB.83.224109},
  url = {https://link.aps.org/doi/10.1103/PhysRevB.83.224109}
}

@article{Lim2021ADiffraction,
    title = {{A convolutional neural network for defect classification in Bragg coherent X-ray diffraction}},
    year = {2021},
    journal = {npj Computational Materials},
    author = {Lim, Bruce and Bellec, Ewen and Dupraz, Maxime and Leake, Steven and Resta, Andrea and Coati, Alessandro and Sprung, Michael and Almog, Ehud and Rabkin, Eugen and Schulli, Tobias and Richard, Marie Ingrid},
    number = {1},
    month = {12},
    volume = {7},
    publisher = {Nature Research},
    doi = {10.1038/s41524-021-00583-9},
    issn = {20573960},
    arxivId = {2106.16179}
}

@article{Marchesini2007ARetrieval,
    title = {{A unified evaluation of iterative projection algorithms for phase retrieval}},
    year = {2007},
    journal = {Review of Scientific Instruments},
    author = {Marchesini, S.},
    number = {1},
    volume = {78},
    doi = {10.1063/1.2403783},
    issn = {00346748},
    pmid = {17503899},
    arxivId = {physics/0603201}
}

@article{Yao2022AutoPhaseNN:Imaging,
    title = {{AutoPhaseNN: unsupervised physics-aware deep learning of 3D nanoscale Bragg coherent diffraction imaging}},
    year = {2022},
    journal = {npj Computational Materials},
    author = {Yao, Yudong and Chan, Henry and Sankaranarayanan, Subramanian and Balaprakash, Prasanna and Harder, Ross J. and Cherukara, Mathew J.},
    number = {1},
    month = {12},
    volume = {8},
    publisher = {Nature Research},
    doi = {10.1038/s41524-022-00803-w},
    issn = {20573960},
    arxivId = {2109.14053}
}

@misc{Robinson2009CoherentNanoscale,
    title = {{Coherent X-ray diffraction imaging of strain at the nanoscale}},
    year = {2009},
    booktitle = {Nature Materials},
    author = {Robinson, Ian and Harder, Ross},
    number = {4},
    pages = {291--298},
    volume = {8},
    publisher = {Nature Publishing Group},
    doi = {10.1038/nmat2400},
    issn = {14764660}
}

@article{FreeLLK,
   author = {Vincent Favre-Nicolin and Steven Leake and Yuriy Chushkin},
   doi = {10.1038/s41598-020-57561-2},
   issn = {20452322},
   issue = {1},
   journal = {Scientific Reports},
   month = {12},
   pmid = {32060293},
   publisher = {Nature Research},
   title = {Free log-likelihood as an unbiased metric for coherent diffraction imaging},
   volume = {10},
   year = {2020}
}

@article{Favre-Nicolin2020PyNX:Operators,
    title = {{PyNX: High-performance computing toolkit for coherent X-ray imaging based on operators}},
    year = {2020},
    journal = {Journal of Applied Crystallography},
    author = {Favre-Nicolin, Vincent and Girard, Gaétan and Leake, Steven and Carnis, Jerome and Chushkin, Yuriy and Kieffer, Jerome and Paleo, Pierre and Richard, Marie Ingrid},
    month = {10},
    pages = {1404--1413},
    volume = {53},
    publisher = {International Union of Crystallography},
    doi = {10.1107/S1600576720010985},
    issn = {16005767},
    arxivId = {2008.11511}
}

@article{Cherukara2018Real-timeNetworks,
    title = {{Real-time coherent diffraction inversion using deep generative networks}},
    year = {2018},
    journal = {Scientific Reports},
    author = {Cherukara, Mathew J. and Nashed, Youssef S.G. and Harder, Ross J.},
    number = {1},
    month = {12},
    volume = {8},
    publisher = {Nature Publishing Group},
    doi = {10.1038/s41598-018-34525-1},
    issn = {20452322},
    pmid = {30410034}
}

@article{Zhao:te5119,
author = "Zhao, Jiangtao and Vartanyants, Ivan A. and Zhang, Fucai",
title = "{Bragg coherent modulation imaging for highly strained nanocrystals: a numerical study}",
journal = "Journal of Applied Crystallography",
year = "2023",
volume = "56",
number = "5",
pages = "1528--1536",
month = "Oct",
doi = {10.1107/S1600576723007720},
url = {https://doi.org/10.1107/S1600576723007720},
}

@article{Carnis2019TowardsReconstructions,
    title = {{Towards a quantitative determination of strain in Bragg Coherent X-ray Diffraction Imaging: artefacts and sign convention in reconstructions}},
    year = {2019},
    journal = {Scientific Reports},
    author = {Carnis, Jérôme and Gao, Lu and Labat, Stéphane and Kim, Young Yong and Hofmann, Jan P. and Leake, Steven J. and Sch{\"{u}}lli, Tobias U. and Hensen, Emiel J.M. and Thomas, Olivier and Richard, Marie Ingrid},
    number = {1},
    month = {12},
    volume = {9},
    publisher = {Nature Research},
    doi = {10.1038/s41598-019-53774-2},
    issn = {20452322},
    pmid = {31758040}
}

@article{minkevich_inversion_2007,
	title = {Inversion of the diffraction pattern from an inhomogeneously strained crystal using an iterative algorithm},
	volume = {76},
	url = {http://link.aps.org/doi/10.1103/PhysRevB.76.104106},
	doi = {10.1103/PhysRevB.76.104106},
	number = {10},
	urldate = {2013-10-15},
	journal = {Phys. Rev. B},
	author = {Minkevich, A. A. and Gailhanou, M. and Micha, J.-S. and Charlet, B. and Chamard, V. and Thomas, O.},
	month = sep,
	year = {2007},
	pages = {104106},
}

@article{clark_high-resolution_2012,
	title = {High-resolution three-dimensional partially coherent diffraction imaging},
	volume = {3},
	copyright = {2012 Nature Publishing Group},
	issn = {2041-1723},
	url = {https://www.nature.com/articles/ncomms1994},
	doi = {10.1038/ncomms1994},
	language = {en},
	urldate = {2019-02-03},
	journal = {Nature Communications},
	author = {Clark, J. N. and Huang, X. and Harder, R. and Robinson, I. K.},
	month = aug,
	year = {2012},
	pages = {993},
}

@article{Minkevich2008,
   author = {A. A. Minkevich and T. Baumbach and M. Gailhanou and O. Thomas},
   doi = {10.1103/PhysRevB.78.174110},
   issn = {10980121},
   issue = {17},
   journal = {Physical Review B - Condensed Matter and Materials Physics},
   month = {11},
   title = {Applicability of an iterative inversion algorithm to the diffraction patterns from inhomogeneously strained crystals},
   volume = {78},
   year = {2008},
}

@article{Ulvestad2017_Guided,
  author    = {Ulvestad, Andrew and Nashed, Youssef and Beutier, Guillaume and Verdier, Michel and Hruszkewycz, Stephan O. and Dupraz, Matthieu},
  title     = {Identifying Defects with Guided Algorithms in Bragg Coherent Diffractive Imaging},
  journal   = {Scientific Reports},
  year      = {2017},
  volume    = {7},
  number    = {1},
  pages     = {9920},
  doi       = {10.1038/s41598-017-09582-7},
  pmid      = {28855571},
  pmcid     = {PMC5577107},
  publisher = {Nature Publishing Group},
  month     = aug,
}

@article{maddali_differentiable_2022,
	title = {A differentiable forward model for the concurrent, multi-peak Bragg coherent x-ray diffraction imaging problem},
	url = {http://arxiv.org/abs/2208.00970},
	author = {Maddali, S. and Frazer, T. D. and Delegan, N. and Harmon, K. J. and Sullivan, S. E. and Allain, M. and Cha, W. and Dibos, A. and Poudyal, I. and Kandel, S. and Nashed, Y. S. G. and Heremans, F. J. and You, H. and Cao, Y. and Hruszkewycz, S. O.},
	date = {2022-08-01},
	eprinttype = {arxiv},
	eprint = {2208.00970},
}

@misc{data4,
  author       = {Bellec, E. and Leake, S. and Richard, M.-I. and Zhao, J.},
  title        = {Bragg Coherent Modulation Imaging for Highly Strained Nanocrystal [Dataset]},
  year         = {2027},
  publisher    = {European Synchrotron Radiation Facility},
  doi          = {10.15151/ESRF-ES-1507879918},
  url          = {https://doi.org/10.15151/ESRF-ES-1507879918},
  note         = {[Dataset]}
}

@software{jax2018github,
  author = {James Bradbury and Roy Frostig and Peter Hawkins and Matthew James Johnson and
            Yash Katariya and Chris Leary and Dougal Maclaurin and George Necula and
            Adam Paszke and Jake Vander{P}las and Skye Wanderman-{M}ilne and Qiao Zhang},
  title = {{JAX}: composable transformations of {P}ython+{N}um{P}y programs},
  url = {http://github.com/jax-ml/jax},
  version = {0.4.0},  
  year = {2018},
}

@article{cdi_noise2017,
author = {Jahn, T. and Wilke, R. N. and Chushkin, Y. and Salditt, T.},
title = {How many photons are needed to reconstruct random objects in coherent X-ray diffractive imaging?},
journal = {Acta Crystallographica Section A},
volume = {73},
number = {1},
pages = {19-29},
doi = {https://doi.org/10.1107/S2053273316015114},
url = {https://onlinelibrary.wiley.com/doi/abs/10.1107/S2053273316015114},
eprint = {https://onlinelibrary.wiley.com/doi/pdf/10.1107/S2053273316015114},
year = {2017}
}

@article{Favre-Nicolin2020,
   author = {Vincent Favre-Nicolin and Steven Leake and Yuriy Chushkin},
   doi = {10.1038/s41598-020-57561-2},
   issn = {20452322},
   issue = {1},
   journal = {Scientific Reports},
   month = {12},
   pmid = {32060293},
   publisher = {Nature Research},
   title = {Free log-likelihood as an unbiased metric for coherent diffraction imaging},
   volume = {10},
   year = {2020}
}

@article{PhysRevLett.87.195505,
  title = {Reconstruction of the Shapes of Gold Nanocrystals Using Coherent X-Ray Diffraction},
  author = {Robinson, I. K. and Vartanyants, I. A. and Williams, G. J. and Pfeifer, M. A. and Pitney, J. A.},
  journal = {Phys. Rev. Lett.},
  volume = {87},
  issue = {19},
  pages = {195505},
  numpages = {4},
  year = {2001},
  month = {Oct},
  publisher = {American Physical Society},
  doi = {10.1103/PhysRevLett.87.195505},
  url = {https://link.aps.org/doi/10.1103/PhysRevLett.87.195505}
}

@article{Zhao,
author = {Zhao, Jiangtao and Vartanyants, Ivan A. and Zhang, Fucai},
title = {Bragg coherent modulation imaging for highly strained nanocrystals: a numerical study},
journal = {Journal of Applied Crystallography},
volume = {56},
number = {5},
pages = {1528-1536},
doi = {https://doi.org/10.1107/S1600576723007720},
url = {https://onlinelibrary.wiley.com/doi/abs/10.1107/S1600576723007720},
eprint = {https://onlinelibrary.wiley.com/doi/pdf/10.1107/S1600576723007720},
year = {2023}
}

@article{NewtonStrain,
  title = {Phase retrieval of diffraction from highly strained crystals},
  author = {Newton, Marcus C. and Harder, Ross and Huang, Xiaojing and Xiong, Gang and Robinson, Ian K.},
  journal = {Phys. Rev. B},
  volume = {82},
  issue = {16},
  pages = {165436},
  numpages = {6},
  year = {2010},
  month = {Oct},
  publisher = {American Physical Society},
  doi = {10.1103/PhysRevB.82.165436},
  url = {https://link.aps.org/doi/10.1103/PhysRevB.82.165436}
}

@article{Luke_2005,
doi = {10.1088/0266-5611/21/1/004},
url = {https://doi.org/10.1088/0266-5611/21/1/004},
year = {2004},
month = {nov},
publisher = {},
volume = {21},
number = {1},
pages = {37},
author = {D Russell Luke},
title = {Relaxed averaged alternating reflections for diffraction imaging},
journal = {Inverse Problems}
}

@article{Masto:yr5131,
author = "Masto, Matteo and Favre-Nicolin, Vincent and Leake, Steven and Sch{\"{u}}lli, Tobias and Richard, Marie-Ingrid and Bellec, Ewen",
title = "{Patching-based deep-learning model for the inpainting of Bragg coherent diffraction patterns affected by detector gaps}",
journal = "Journal of Applied Crystallography",
year = "2024",
volume = "57",
number = "4",
pages = "966--974",
month = "Aug",
doi = {10.1107/S1600576724004163},
url = {https://doi.org/10.1107/S1600576724004163}
}

@article{Vartanyants_2001,
doi = {10.1088/0953-8984/13/47/305},
url = {https://doi.org/10.1088/0953-8984/13/47/305},
year = {2001},
month = {nov},
publisher = {},
volume = {13},
number = {47},
pages = {10593},
author = {I A Vartanyants and I K Robinson},
title = {Partial coherence effects on the imaging of small crystals using coherent x-ray diffraction},
journal = {Journal of Physics: Condensed Matter}
}

@article{stoclet2025structural,
  title={Structural dependency of polymer dynamics by means of small-angle X-ray photon correlation spectroscopy and wide-angle X-ray scattering on the D2AM beamline},
  author={Stoclet, Gr{\'e}gory and Schwaller, Duncan and Garlet, Romain and Livet, Fr{\'e}d{\'e}ric and Chahine, Gilbert A and Blanc, Nils and Dupraz, Maxime},
  journal={Synchrotron Radiation},
  volume={32},
  number={3},
  pages={649--660},
  year={2025},
  publisher={International Union of Crystallography}
}

@article{met9030352,
AUTHOR = {Chahine, Gilbert André and Blanc, Nils and Arnaud, Stephan and De Geuser, Frédéric and Guinebretière, René and Boudet, Nathalie},
TITLE = {Advanced Non-Destructive in Situ Characterization of Metals with the French Collaborating Research Group D2AM/BM02 Beamline at the European Synchrotron Radiation Facility},
JOURNAL = {Metals},
VOLUME = {9},
YEAR = {2019},
NUMBER = {3},
ARTICLE-NUMBER = {352},
URL = {https://www.mdpi.com/2075-4701/9/3/352},
ISSN = {2075-4701},
DOI = {10.3390/met9030352}
}

@article{Richard:te5091,
author = "Richard, M.-I. and Labat, S. and Dupraz, M. and Li, N. and Bellec, E. and Boesecke, P. and Djazouli, H. and Eymery, J. and Thomas, O. and Sch{\"{u}}lli, T. U. and Santala, M. K. and Leake, S. J.",
title = "{Bragg coherent diffraction imaging of single 20nm Pt particles at the ID01-EBS beamline of ESRF}",
journal = "Journal of Applied Crystallography",
year = "2022",
volume = "55",
number = "3",
pages = "621--625",
month = "Jun",
doi = {10.1107/S1600576722002886},
url = {https://doi.org/10.1107/S1600576722002886},
}

@Article{Atlan2023,
author={Atlan, Cl{\~A}{\textcopyright}ment
and Chatelier, Corentin
and Martens, Isaac
and Dupraz, Maxime
and Viola, Arnaud
and Li, Ni
and Gao, Lu
and Leake, Steven J.
and Schülli, Tobias U.
and Eymery, Jo{\~A}«l
and Maillard, Fr{\~A}{\textcopyright}d{\~A}{\textcopyright}ric
and Richard, Marie-Ingrid},
title={Imaging the strain evolution of a platinum nanoparticle under electrochemical control},
journal={Nature Materials},
year={2023},
month={Jun},
day={01},
volume={22},
number={6},
pages={754-761},
issn={1476-4660},
doi={10.1038/s41563-023-01528-x},
url={https://doi.org/10.1038/s41563-023-01528-x}
}

@misc{kingma2017adam,
      title={Adam: A Method for Stochastic Optimization}, 
      author={Diederik P. Kingma and Jimmy Ba},
      year={2017},
      eprint={1412.6980},
      archivePrefix={arXiv},
      primaryClass={cs.LG}
}

@article{Godard:te5072,
author = "Godard, Pierre",
title = "{On the use of the scattering amplitude in coherent X-ray Bragg diffraction imaging}",
journal = "Journal of Applied Crystallography",
year = "2021",
volume = "54",
number = "3",
pages = "797--802",
month = "Jun",
doi = {10.1107/S1600576721003113},
url = {https://doi.org/10.1107/S1600576721003113},
}

@article{Wu:cw5029,
author = "Wu, Longlong and Juhas, Pavol and Yoo, Shinjae and Robinson, Ian",
title = "{Complex imaging of phase domains by deep neural networks}",
journal = "IUCrJ",
year = "2021",
volume = "8",
number = "1",
pages = "12--21",
month = "Jan",
doi = {10.1107/S2052252520013780},
url = {https://doi.org/10.1107/S2052252520013780},
}

@article{Yang2013,
   author = {Wenge Yang and Xiaojing Huang and Ross Harder and Jesse N. Clark and Ian K. Robinson and Ho Kwang Mao},
   doi = {10.1038/ncomms2661},
   issn = {20411723},
   journal = {Nature Communications},
   title = {Coherent diffraction imaging of nanoscale strain evolution in a single crystal under high pressure},
   volume = {4},
   year = {2013}
}

@article{Maddali2023,
   author = {Siddharth Maddali and Travis D. Frazer and Nazar Delegan and Katherine J. Harmon and Sean E. Sullivan and Marc Allain and Wonsuk Cha and Alan Dibos and Ishwor Poudyal and Saugat Kandel and Youssef S.G. Nashed and F. Joseph Heremans and Hoydoo You and Yue Cao and Stephan O. Hruszkewycz},
   doi = {10.1038/s41524-023-01022-7},
   issn = {20573960},
   issue = {1},
   journal = {npj Computational Materials},
   month = {12},
   publisher = {Nature Research},
   title = {Concurrent multi-peak Bragg coherent x-ray diffraction imaging of 3D nanocrystal lattice displacement via global optimization},
   volume = {9},
   year = {2023}
}

@misc{data_mouad,
  author       = {Bellec, E. and Khater, P. and Richard, M.-I. and Bouita, M. and Leake, S. and Sarrazin, T.},
  title        = {BCDI measurement of a Pt particle on an YSZ substrate (Version 1) [Dataset]},
  year         = {2026},
  publisher    = {European Synchrotron Radiation Facility},
  doi          = {10.15151/ESRF-DC-2538627445},
  url          = {https://doi.org/10.15151/ESRF-DC-2538627445},
  note         = {Includes data from scan\_266}
}

@misc{data_bm02,
  author       = {Bellec, E. and Richard, M. I.},
  title        = {Bragg Coherent Diffraction Imaging of PT nanoparticle with dislocation (Version 1) },
  year         = {2026},
  publisher    = {European Synchrotron Radiation Facility},
  doi          = {10.15151/ESRF-DC-2543797486},
  url          = {https://doi.org/10.15151/ESRF-DC-2543797486},
}

@article{Fienup:78,
author = {J. R. Fienup},
journal = {Opt. Lett.},
number = {1},
pages = {27--29},
publisher = {Optica Publishing Group},
title = {Reconstruction of an object from the modulus of its Fourier transform},
volume = {3},
month = {Jul},
year = {1978},
url = {https://opg.optica.org/ol/abstract.cfm?URI=ol-3-1-27},
doi = {10.1364/OL.3.000027},
}

@article{Gerchberg1972APA,
  title={A practical algorithm for the determination of phase from image and diffraction plane pictures},
  author={R. W. Gerchberg},
  journal={Optik},
  year={1972},
  volume={35},
  pages={237-246},
  url={https://api.semanticscholar.org/CorpusID:55691159}
}

@article{PhysRevMaterials.4.013801,
  title = {Nanoscale imaging of the full strain tensor of specific dislocations extracted from a bulk sample},
  author = {Hofmann, Felix and Phillips, Nicholas W. and Das, Suchandrima and Karamched, Phani and Hughes, Gareth M. and Douglas, James O. and Cha, Wonsuk and Liu, Wenjun},
  journal = {Phys. Rev. Mater.},
  volume = {4},
  issue = {1},
  pages = {013801},
  numpages = {12},
  year = {2020},
  month = {Jan},
  publisher = {American Physical Society},
  doi = {10.1103/PhysRevMaterials.4.013801},
  url = {https://link.aps.org/doi/10.1103/PhysRevMaterials.4.013801}
}

@article{dzhigaev2021three,
  title={Three-dimensional coherent x-ray diffraction imaging of ferroelastic domains in single CsPbBr3 perovskite nanoparticles},
  author={Dzhigaev, D and Zhang, Z and Mar{\c{c}}al, LAB and Sala, S and Bj{\"o}rling, A and Mikkelsen, A and Wallentin, J},
  journal={New Journal of Physics},
  volume={23},
  number={6},
  pages={063035},
  year={2021},
  publisher={IOP Publishing}
}

@article{diao2020evolution,
  title={Evolution of ferroelastic domain walls during phase transitions in barium titanate nanoparticles},
  author={Diao, Jiecheng and Shi, Xiaowen and Assefa, Tadesse A and Wu, Longlong and Suzana, Ana F and Nunes, Daniel S and Batey, Darren and Cipiccia, Silvia and Rau, Christoph and Harder, Ross J and others},
  journal={Physical Review Materials},
  volume={4},
  number={10},
  pages={106001},
  year={2020},
  publisher={APS}
}

@article{zakaria2026strain,
  title={Strain Evolution in Nanocrystals under High Pressure Tracked with Bragg Coherent X-ray Diffraction Imaging},
  author={Zakaria, Abdelrahman and Yehya, Sarah and Fertey, Pierre and Wehinger, Björn and Leake, Steven and Levi, Mor and Cornelius, Thomas W and Thomas, Olivier and Mezouar, Mohamed and Rabkin, Eugen and others},
  journal={Nano Letters},
  volume={26},
  number={14},
  pages={4807--4814},
  year={2026},
  publisher={ACS Publications}
}

@article{olson2025kinetics,
  title={Kinetics of Hydrogen Absorption in Individual $\alpha$-Phase Palladium Nanoparticles},
  author={Olson, Kyle J and Viola, Arnaud and Bellec, Ewen and Atlan, Cl{\'e}ment and Chatelier, Corentin and Grimes, Michael and Gilles, Bruno and Schulli, Tobias U and Leake, Steven J and Vandichel, Matthias and others},
  journal={ACS Applied Energy Materials},
  volume={8},
  number={14},
  pages={10256--10264},
  year={2025},
  publisher={ACS Publications}
}

@article{atlan2025probing,
  title={Probing Strain in Individual Palladium Nanocrystals during Electrochemically Induced Phase Transitions},
  author={Atlan, Cl{\'e}ment and Chatelier, Corentin and Ngoipala, Apinya and Olson, Kyle and Viola, Arnaud and Bellec, Ewen and Grimes, Michael and Gilles, Bruno and Qamar, Minaam and Mrovec, Matous and others},
  journal={Journal of the American Chemical Society},
  volume={147},
  number={29},
  pages={25417--25428},
  year={2025},
  publisher={ACS Publications}
}

@article{grimes2024capturing,
  title={Capturing catalyst strain dynamics during operando CO oxidation},
  author={Grimes, Michael and Atlan, Cl{\'e}ment and Chatelier, Corentin and Bellec, Ewen and Olson, Kyle and Simonne, David and Levi, Mor and Schulli, Tobias U and Leake, Steven J and Rabkin, Eugen and others},
  journal={ACS nano},
  volume={18},
  number={30},
  pages={19608--19617},
  year={2024},
  publisher={ACS Publications}
}

@article{chatelier2024unveiling,
  title={Unveiling core--shell structure formation in a ni3fe nanoparticle with in situ multi-bragg coherent diffraction imaging},
  author={Chatelier, Corentin and Atlan, Cl{\'e}ment and Dupraz, Maxime and Leake, Steven and Li, Ni and Schulli, Tobias U and Levi, Mor and Rabkin, Eugen and Favre, Luc and Labat, St{\'e}phane and others},
  journal={ACS nano},
  volume={18},
  number={21},
  pages={13517--13527},
  year={2024},
  publisher={ACS Publications}
}

@Article{Vicente2026,
author={Vicente, Rafael Alcides
and Yukuhiro, Victor Yoiti
and Atlan, Cl{\'e}ment
and Bellec, Ewen
and Zhao, Jiangtao
and Richard, Marie-Ingrid
and Neckel, Itamar Tomio
and Fern{\'a}ndez, Pablo Sebasti{\'a}n},
title={Unveiling the Effects of Adsorbates on Pt Nanoparticle Crystal Structure during an Electrocatalytic Reaction},
journal={ACS Catalysis},
year={2026},
month={Jun},
day={05},
publisher={American Chemical Society},
volume={16},
number={11},
pages={10163-10173},
doi={10.1021/acscatal.6c00873},
url={https://doi.org/10.1021/acscatal.6c00873}
}

@article{yang2025low,
  title={Low-temperature state in strontium titanate microcrystals using in situ multireflection Bragg coherent x-ray diffraction imaging},
  author={Yang, David and Suzana, Ana F and Wu, Longlong and Ha, Sung Soo and Choi, Sungwook and Ngo, Hieu Minh and Nawaz, Muhammad Mahmood and Kim, Hyunjung and Liu, Jialun and Treuherz, Daniel and others},
  journal={Physical Review B},
  volume={111},
  number={2},
  pages={024107},
  year={2025},
  publisher={APS}
}

@article{comby2025combining,
  title={Combining nanoindentation and Bragg coherent diffraction imaging to investigate small-scale plasticity},
  author={Comby-Dassonneville, Sol{\`e}ne and Beutier, Guillaume and Parry, Guillaume and Leake, Steven and Charlot, Fr{\'e}d{\'e}ric and Volpi, Fabien and Verdier, Marc},
  journal={Materialia},
  volume={39},
  pages={102376},
  year={2025},
  publisher={Elsevier}
}

@article{labat2015inversion,
  title={Inversion domain boundaries in GaN wires revealed by coherent Bragg imaging},
  author={Labat, St{\'e}phane and Richard, Marie-Ingrid and Dupraz, Maxime and Gailhanou, Marc and Beutier, Guillaume and Verdier, Marc and Mastropietro, Francesca and Cornelius, Thomas W and Schulli, Tobias U and Eymery, Joel and others},
  journal={ACS nano},
  volume={9},
  number={9},
  pages={9210--9216},
  year={2015},
  publisher={ACS Publications}
}

@article{wang2020algorithm,
  title={An algorithm for Bragg coherent x-ray diffractive imaging of highly strained nanocrystals},
  author={Wang, Ziyi and Gorobtsov, Oleg and Singer, Andrej},
  journal={New Journal of Physics},
  volume={22},
  number={1},
  pages={013021},
  year={2020},
  publisher={IOP Publishing}
}

@article{zhao2025bragg,
  title={Bragg Coherent Modulation Imaging of Highly Strained Nanocrystals},
  author={Zhao, Jiangtao and Bellec, Ewen and Richard, Marie-Ingrid and Pithan, Linus and Vartanyants, Ivan A and Zhang, Fucai and Sch{\"u}lli, Tobias and Leake, Steven J},
  journal={Physical Review Letters},
  volume={135},
  number={25},
  pages={256101},
  year={2025},
  publisher={APS}
}

@article{hruszkewycz2017high,
  title={High-resolution three-dimensional structural microscopy by single-angle Bragg ptychography},
  author={Hruszkewycz, Stephan O and Allain, Marc and Holt, Martin V and Murray, Conal E and Holt, Judson R and Fuoss, Paul H and Chamard, Virginie},
  journal={Nature materials},
  volume={16},
  number={2},
  pages={244--251},
  year={2017},
  publisher={Nature Publishing Group UK London}
}

@article{li2021revealing,
  title={Revealing nano-scale lattice distortions in implanted material with 3D Bragg ptychography},
  author={Li, Peng and Phillips, Nicholas W and Leake, Steven and Allain, Marc and Hofmann, Felix and Chamard, Virginie},
  journal={Nature communications},
  volume={12},
  number={1},
  pages={7059},
  year={2021},
  publisher={Nature Publishing Group UK London}
}

@article{takahashi2013bragg,
  title={Bragg x-ray ptychography of a silicon crystal: Visualization of the dislocation strain field and the production of a vortex beam},
  author={Takahashi, Yukio and Suzuki, Akihiro and Furutaku, Shin and Yamauchi, Kazuto and Kohmura, Yoshiki and Ishikawa, Tetsuya},
  journal={Physical Review B—Condensed Matter and Materials Physics},
  volume={87},
  number={12},
  pages={121201},
  year={2013},
  publisher={APS}
}

@article{pelzer2021removal,
  title={Removal of spurious data in Bragg coherent diffraction imaging: an algorithm for automated data preprocessing},
  author={Pelzer, Kenley and Schwarz, Nicholas and Harder, Ross},
  journal={Applied Crystallography},
  volume={54},
  number={2},
  pages={523--532},
  year={2021},
  publisher={International Union of Crystallography}
}

@article{bellec2026machine,
  title={Machine learning assisted masking of parasitic signals in Bragg coherent diffraction imaging},
  author={Bellec, Ewen and Leake, Steven J and Levi, Mor and Rabkin, Eugen and Sch{\"u}lli, Tobias U and Richard, M-I},
  journal={Synchrotron Radiation},
  volume={33},
  number={2},
  year={2026},
  publisher={International Union of Crystallography}
}

@article{Masto2026,
  author = {Masto, Matteo and Favre-Nicolin, Vincent and Leake, Steven and Schülli, Tobias and Richard, Marie-Ingrid and Atlan, Clement and Bellec, Ewen},
  title = {Phase retrieval of highly strained Bragg coherent diffraction patterns using supervised convolutional neural network},
  journal = {npj Computational Materials},
  year = {2026},
  volume = {12},
  number = {1},
  pages = {164},
  doi = {10.1038/s41524-026-02017-w},
  url = {https://doi.org/10.1038/s41524-026-02017-w},
  issn = {2057-3960}
}

@article{Wu2025,
  author = {Wu, Longlong and Yang, David and Wang, Wei and Yoo, Shinjae and Harder, Ross J. and Cha, Wonsuk and Li, Aiguo and Robinson, Ian K.},
  title = {Unveiling nano-scale crystal deformation using coherent X-ray dynamical diffraction},
  journal = {npj Computational Materials},
  year = {2025},
  volume = {11},
  number = {1},
  pages = {379},
  month = dec,
  doi = {10.1038/s41524-025-01858-1},
  url = {https://doi.org/10.1038/s41524-025-01858-1},
  issn = {2057-3960}
}

@article{Jurling:14,
author = {Alden S. Jurling and James R. Fienup},
journal = {J. Opt. Soc. Am. A},
number = {7},
pages = {1348--1359},
publisher = {Optica Publishing Group},
title = {Applications of algorithmic differentiation to phase retrieval algorithms},
volume = {31},
month = {Jul},
year = {2014},
url = {https://opg.optica.org/josaa/abstract.cfm?URI=josaa-31-7-1348},
doi = {10.1364/JOSAA.31.001348},
}

@article{kandel2019using,
  title={Using automatic differentiation as a general framework for ptychographic reconstruction},
  author={Kandel, Saugat and Maddali, S and Allain, Marc and Hruszkewycz, Stephan O and Jacobsen, Chris and Nashed, Youssef SG},
  journal={Optics express},
  volume={27},
  number={13},
  pages={18653--18672},
  year={2019},
  publisher={Optical Society of America}
}

@article{frisch2023unraveling,
  title={Unraveling the synergistic effects of Cu-Ag tandem catalysts during electrochemical CO2 reduction using nanofocused X-ray probes},
  author={Frisch, Marvin L and Wu, Longfei and Atlan, Cl{\'e}ment and Ren, Zhe and Han, Madeleine and Tucoulou, R{\'e}mi and Liang, Liang and Lu, Jiasheng and Guo, An and Nong, Hong Nhan and others},
  journal={Nature Communications},
  volume={14},
  number={1},
  pages={7833},
  year={2023},
  publisher={Nature Publishing Group UK London}
}

@Article{Björling2023,
author={Bj{\"o}rling, Alexander
and Mar{\c{c}}al, Lucas A. B.
and Ar{\'a}n-Ais, Rosa M.
and Solla-Gull{\'o}n, Jose},
title={Chemical Limits on X-ray Nanobeam Studies in Water},
journal={The Journal of Physical Chemistry C},
year={2023},
month={Jul},
day={20},
publisher={American Chemical Society},
volume={127},
number={28},
pages={13877-13885},
issn={1932-7447},
doi={10.1021/acs.jpcc.3c02432},
url={https://doi.org/10.1021/acs.jpcc.3c02432}
}

@article{SIMONNE2026116940,
title = {X-ray irradiation induced strain relaxation of dewetted Ni particles on modified Si substrate},
journal = {Scripta Materialia},
volume = {270},
pages = {116940},
year = {2026},
issn = {1359-6462},
doi = {https://doi.org/10.1016/j.scriptamat.2025.116940},
url = {https://www.sciencedirect.com/science/article/pii/S1359646225004026},
author = {David Simonne and Riley Hultquist and Jiangtao Zhao and Andrea Resta and Ericmoore Jossou}}

@Article{Jousseaume2023,
author={Jousseaume, Thibaut
and Colin, Jean-Fran{\c{c}}ois
and Chandesris, Marion
and Lyonnard, Sandrine
and Tardif, Samuel},
title={How Beam Damage Can Skew Synchrotron Operando Studies of Batteries},
journal={ACS Energy Letters},
year={2023},
month={Aug},
day={11},
publisher={American Chemical Society},
volume={8},
number={8},
pages={3323-3329},
doi={10.1021/acsenergylett.3c00815},
url={https://doi.org/10.1021/acsenergylett.3c00815}
}

@Article{Black2024,
author={Black, Ashley P.
and Escudero, Carlos
and Fauth, Fran{\c{c}}ois
and Fehse, Marcus
and Agostini, Giovanni
and Reynaud, Marine
and Houdeville, Raphaelle G.
and Chatzogiannakis, Dimitrios
and Orive, Joseba
and Ramo-Irurre, Alejandro
and Casas-Cabanas, Montse
and Palacin, M. Rosa},
title={Beam Effects in Synchrotron Radiation Operando Characterization of Battery Materials: X-Ray Diffraction and Absorption Study of LiNi0.33Mn0.33Co0.33O2 and LiFePO4 Electrodes},
journal={Chemistry of Materials},
year={2024},
month={Jun},
day={11},
publisher={American Chemical Society},
volume={36},
number={11},
pages={5596-5610},
issn={0897-4756},
doi={10.1021/acs.chemmater.4c00597},
url={https://doi.org/10.1021/acs.chemmater.4c00597}
}

\end{document}